\documentclass[useAMS,usenatbib]{mn2e}
\usepackage{psfig}
\usepackage{graphicx}
\usepackage{graphics}
\usepackage{epsfig}
\usepackage{epsf}
\newcommand{\mnras}{\,{\rm MNRAS}}
\newcommand{\apj}{\,{\rm ApJ}}
\newcommand{\apjl}{\,{\rm ApJL}}

\newcommand{\aap}{\,{\rm A\&Ap}}
\newcommand{\apss}{\,{\rm Ap\&SS}}
\newcommand{\aj}{\,{\rm AJ}}
\newcommand{\prd}{\,{\rm PRvd}}

\newcommand{\ssst}{\scriptscriptstyle}
\newcommand{\E}[1]{\times 10^{#1}}
\newcommand{\etal}{et al.}

\newcommand{\s}{\,{\rm s}}      \newcommand{\ps}{\,{\rm s}^{-1}}
\newcommand{\yr}{\,{\rm yr}}    
\newcommand{\cm}{\,{\rm cm}}    \newcommand{\km}{\,{\rm km}}
\newcommand{\parsec}{\,{\rm pc}}

\newcommand{\no}{n_{\ssst 0}}   \newcommand{\ti}{t_{\rm i}}
\newcommand{\Rc}{R_{\rm c}}     \newcommand{\Rs}{R_{\rm s}}
\newcommand{\RSNR}{R_{\ssst\rm SNR}}    \newcommand{\vSNR}{v_{\ssst\rm SNR}}
\newcommand{\ESNR}{E_{\ssst\rm SNR}}
        \newcommand{\mH}{m_{\ssst\rm H}}

\newcommand{\gray}{$\gamma$-ray}    \newcommand{\grays}{$\gamma$-rays}

\newcommand{\HESS}{{\rm H.E.S.S.}}  \newcommand{\Fermi}{{\sl Fermi}}

\title[\gray\ Spectra of SNR W28]
{
\grays\ from molecular clouds illuminated by accumulated diffusive protons from supernova remnant W28}

\author[H. Li et al.]
{Hui Li$^{1}$
~and Yang Chen$^{1,2}$\thanks{E-mail: ygchen@nju.edu.cn}
\\
$^{1}$Department of Astronomy, Nanjing University, Nanjing 210093, P.\ R.\ China\\
$^{2}$Key Laboratory of Modern Astronomy and Astrophysics, Nanjing University, Ministry of Education, Nanjing 210093, China\\
}

\begin{document}

\date{Accepted . Received ; in original form}

\pagerange{\pageref{firstpage}--\pageref{lastpage}} \pubyear{2009}

\maketitle

\label{firstpage}

\begin{abstract}
W28 is one of the archetype supernova remnants (SNRs) interacting
with molecular clouds. \HESS\ observation found four TeV sources
which are coincident with the molecular clouds (MCs) around W28, but
\Fermi\ LAT detected no prominent GeV counterparts for two of them.
An accumulative diffusion model is established in this Letter and
the energetic protons colliding the nearby MCs are considered to be
an accumulation of the diffusive protons escaping from the shock
front throughout the history of the SNR expansion.
We have fitted the \gray\ spectra of the four sources and naturally
explained the GeV spectral break of the northeastern source (source
N) and the nonsignificant GeV emission of the southern sources A and C.
The distances of sources A and C from the SNR centre are found to be
much larger than those of sources N and B, which may be the basic
reason for the faint GeV \grays\ of the two former sources.
\end{abstract}


\section{INTRODUCTION}\label{sec:intro}
Cosmic rays (CRs) below the ``knee" in the Galaxy are commonly
believed to be accelerated at the shock of supernova remnants
(SNRs). However, whether the $\gamma$-rays from SNRs are of hadronic
or leptonic origin is still in hot debate. Although evidences for
$\gamma$-ray emission arising from molecular clouds (MCs) have been
found in two SNRs, IC~443 (Acciari \etal\ 2009) and W28 (Aharonian
\etal\ 2008), it is not easy yet to confirm the emission of the
accelerated protons, since the competing lepton processes can also
account for the high energy radiation. Fortunately, emission from
MCs illuminated by CRs from nearby SNRs (Aharonian \& Atoyan 1996;
Gabici \etal\,2009) can give us opportunities not only to
distinguish the contribution of the hadrons from that of leptons
but also to investigate the escape process of these energetic
protons (Fujita \etal\,2009).
SNR W28, with recent observation in GeV and TeV $\gamma$-rays, is
such an excellent case.

W28 (G6.4$-$0.1) is one of the prototype thermal composite (or mixed-morphology)
SNRs, characterized by centre-filled thermal X-ray and shell-like
radio emission, with large (about $50'\times48'$) apparent
dimensions (Green 2009). It is considered to be an evolved SNR in
the radiative stage, with an age estimate spanning from 35 to 150
kyr (Kaspi \etal\ 1993). SNR W28 has been confirmed to be
interacting with MCs with robust evidences including the 1720 MHz OH
masers, morphological agreement with molecular features, molecular
line broadening, etc.\ (see Jiang \etal\ 2010; and references
therein). The observations of molecular lines place it at a distance
of $\sim2\rm kpc$ (Vel$\acute{\rm a}$zquez \etal\ 2002).

In \grays, $\HESS$ has revealed four TeV sources in the
W28 field positionally coincident with MCs (Aharonian \etal\ 2008):
HESS J1801-233 (denoted as source N in this Letter), located along
the northeastern boundary of W28, and HESS J1800-240A, B, and C
(denoted as sources A, B, and C, respectively, in below), located
$\sim30'$ south of W28. 
Recently, \Fermi\ LAT observation (Abdo \etal\,2010) found GeV
sources 1FGL J1801.3-2322c and 1FGL J1800.5-2359c, which coincide
with sources N and B, respectively. More interestingly, it did not
detect significant GeV counterparts for sources A and C. The LAT
upper limits combined with the \HESS\ data imply a spectral break
between 10GeV and 100GeV. We will show in this Letter that these
\gray\ spatial and spectral properties can be accounted for with an
accumulative diffusion process of CRs which are accelerated at the
dynamically evolving blast shock front.


\section{Accumulative Diffusion Model}\label{sec:model}

CR diffusion models have been developed by Aharonian \& Atoyan
(1996) and Gabici \etal\ (2009) to evaluate the CR spectrum in the
vicinity of SNRs. In their models, both cases of impulsive and
continuous injection are considered on the assumption that the
distance between the SNR and the cloud is significantly larger than
the size of the SNR, so that the injection region can be
approximated as a point-like source. If, however, the cloud is at or
close to the SNR blast shock, such as in the case of W28,
the diffusion distance for protons injected at various time, $\ti$,
and various shock radius, $\Rs$, is not a constant and the
assumption of point-like source injection should be modified.
Therefore, we here establish an accumulative diffusion model
considering that the energetic protons colliding the given MC are a
collection of the diffusive protons escaping from different $\Rs$ as
the SNR expands.
{\em Small distance between the SNR and MC is allowed.}

\begin{figure} 
\centerline { {\hfil\hfil
\psfig{figure=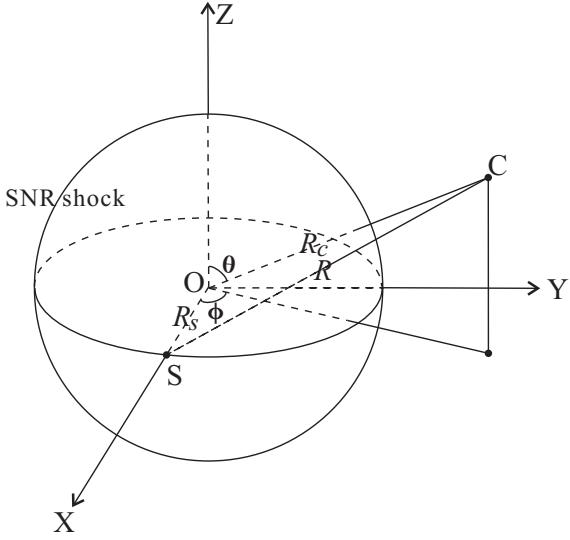,height=2.8in,angle=0}
\hfil\hfil } }
\caption{Sketch for the positional relation between the source point (S) and the field point (C)
 (see text in \S\ref{sec:model}).} \label{fig:W28cartoon}
\end{figure}

The distribution function at an arbitrary field point C (at radius
$\Rc$) of the energetic protons that escape from unit area at an
arbitrary source point S on the spherical shock front surface (see
Figure~\ref{fig:W28cartoon}) is given by (Aharonian \& Atoyan 1996)
\begin{equation}\label{eq:imp-dif}
f(E_{\rm p},R(\ti,\theta,\phi),t_{\rm dif})\approx\frac{1}{4\pi \Rs^2}\frac{dN_0/d\ti}{\pi^{3/2}R_{\rm dif}^3}
\exp\left(-\frac{R^2(\ti,\theta,\phi)}{R_{\rm dif}^2}\right)
\end{equation}
where $R(\ti,\theta,\phi)=\sqrt{R_{\rm s}^2(\ti)+R_{\rm c}^2-2R_{\rm
s}R_{\rm c}\sin\theta\cos\phi}$ is the distance between points S and
C (Figure~\ref{fig:W28cartoon}), $t_{\rm dif}=t_{\rm age}-\ti$ is
the diffusion time after the escape, $R_{\rm dif}\equiv R_{\rm
dif}(E_{\rm p},t_{\rm dif}) =2\sqrt{D(E_{\rm p})t_{\rm dif}}$ is the
diffusion radius, and $dN_0/d\ti\equiv Q_0E_{\rm p}^{-\Gamma}$ is
the total escaping rate at time $\ti$. Here the diffusion
coefficient is assumed to be in the form of $D(E_{\rm
p})=10^{28}\chi(E_{\rm p}/10\rm GeV)^{\delta}\cm^2\s^{-1}$, where
$\chi$ is the correction factor of slow diffusion around the SNR
(Fujita \etal\, 2009) and $\delta\approx0.3$--0.7 (Berezinskii
1990).
The number of protons with energy $E_{\rm p}$ escaping from the SNR
shock front surface at $\ti$ and arriving at the spherical surface
of radius $\Rc$ at $t_{\rm age}$ is \( 4\pi\Rs^2
\int^{2\pi}_0\!\!\!\int^{\pi}_0\!\!\! f(E_{\rm
p},R(\ti,\theta,\phi),t_{\rm dif}) \Rc^2\sin\theta d\theta d\phi \).
Because of spherical symmetry, the distribution function of protons
(escaping from $\Rs(\ti)$) for any point (e.g., point C) on the
spherical surface of radius $\Rc$ is \( \Rs^2
\int^{2\pi}_0\!\!\!\int^{\pi}_0\!\!\! f(E_{\rm
p},R(\ti,\theta,\phi),t_{\rm dif}) \sin\theta d\theta d\phi \).
Therefore,
the distribution function for point C accumulating all the
historical contributions can be obtained as
\[F_{\rm ac}(E_{\rm p},\Rc,t_{\rm age}) = \]
\begin{equation}\label{eq:imp-dif}
\int_0^{t_{\rm age}}\int^{2\pi}_0\int^{\pi}_0 f(E_{\rm
p},R(\ti,\theta,\phi),t_{\rm dif}) \Rs^2\sin\theta\,d\theta\,d\phi\,d\ti.
\end{equation}

For the dynamical evolution of radius of the SNR expanding in the
interstellar (intercloud) medium of density $\rho_0=1.4\mH \no$, we
use the Sedov-Taylor law $R_{\rm s}=(2.026 \ESNR t^2/\rho_{\ssst
0})^{1/5}$ for the adiabatic phase and $R_{\rm s}=(147\epsilon \ESNR
R_{\rm t}^2t^2/4\pi\rho_{\ssst 0})^{1/7}$ for the radiative phase,
where $R_{\rm t}$ is the transition radius from the Sedov phase to
the radiative phase, $\ESNR$ is the supernova explosion energy, and
$\epsilon$ is a factor equal to 0.24 (Blinnikov \etal\ 1982; see
also Lozinskaya 1992).

The total energy of the escaping protons is $W_{\rm p}\equiv\eta
\ESNR=t_{\rm age}\int Q_0E_{\rm p}^{-\Gamma}$ $E_{\rm p}dE_{\rm p}$,
where $\eta$ is the fraction of the explosion energy converted into
protons.

In addition to the escaping protons, we also consider the
contribution of diffuse Galactic protons in the Solar neighborhood
with flux density (see, e.g., Dermer 1986):
\begin{equation}\label{eq:GCRs-dif}
J_{\rm CR}(E_{\rm p})=2.2\left(\frac{E_{\rm p}}{\rm GeV}\right)^{-2.75} \cm^{-2} \s^{-1} \rm GeV^{-1} \rm sr^{-1}.
\end{equation}

In the calculation of the $\gamma$-rays from the nearby MC (of mass
$M_{\rm c}$) due to p-p interaction so as to match the observed GeV
and TeV fluxes, we use the analytic photon emissivity
$dN_{\gamma}/dE_{\gamma}$ developed by Kelner et al.\ (2006). At
high energies,
\begin{equation}
\frac{dN_{\gamma}}{dE_{\gamma}}=cn_{\rm b}\int^{\infty}_{E_{\gamma}}
\sigma_{\rm inel}(E_{\rm p})J_{\rm p}(E_{\rm p})F_{\gamma}\left(\frac{E_{\gamma}}{E_{\rm p}},E_{\rm p}\right)
\frac{dE_{\rm p}}{E_{\rm p}}
\end{equation}
where $n_{\rm b}$ is the average density of target baryons of the
cloud,
$\sigma_{\rm inel}$ is the cross section of inelastic p-p
interactions and function $F_{\gamma}(x,E_{\rm p})$ is defined by
Eq.~(58) in Kelner et al.\ (2006); while at low energies, with the
modified $\delta$-function approximation as proposed in Aharonian \&
Atoyan (2000), the emissivity of $\gamma$-rays is
\begin{equation}
\frac{dN_{\gamma}}{dE_{\gamma}}=2\int^{\infty}_{E_{\rm min}}
\frac{q_{\pi}(E_{\pi})}{\sqrt{E^2_{\pi}-m^2_{\pi}}}dE_{\pi}
\end{equation}
where $E_{\rm min}=E_{\gamma}+m^2_{\pi}/4E_{\gamma}$, and the
production rate of $\pi^0$ mesons is
\begin{equation}
q_{\pi}(E_{\rm p})=\tilde{n}\frac{cn_{\rm H}}{K_{\pi}}\sigma_{\rm inel}
\left(m_{\rm p}+\frac{E_{\pi}}{K_{\pi}}\right)J_{\rm p}\left(m_{\rm p}+\frac{E_{\pi}}{K_{\pi}}\right)
\end{equation}
(see Kelner et al.\ 2006 for the explanations of other parameters).

The contribution of $\gamma$-rays from the secondary leptons that
are created by p-p interaction is negligible, as compared with the
dominant contribution of the protons themselves (Gabici \etal\ 2007;
2009). Because the timescale for the secondaries to escape from the
cloud is shorter than the energy loss time for particle energies
between $\sim100$~MeV and a few 100~TeV, a large amount of the
secondaries produced within clouds of a wide range of parameters
escape without being affected by significant losses (Gabici et al.\
2009). In the molecular gas of density as high as $10^{4}\cm^{-3}$,
the magnetic field could be assumed to be of order $10^2\mu$G
(Crutcher 1999). Extremely high energy ($\ga$ a few 100~TeV)
secondaries radiate all their energy by synchrotron peaked at tens
of~keV before escaping from the cloud; but the X-rays are not in the
scope of this letter.

\section{Application to SNR W28}\label{sec:dif-W28}

\begin{figure*} 
\centerline { {\hfil\hfil
\psfig{figure=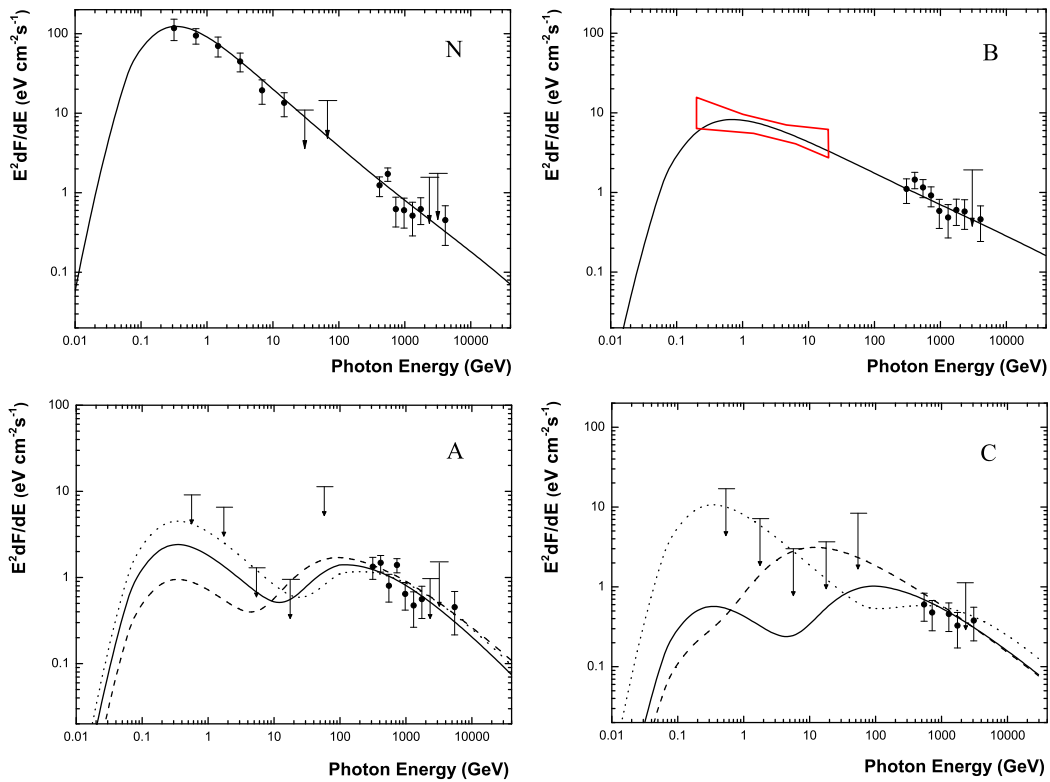,height=4.8in,angle=0} \hfil\hfil } }
\caption{\gray\ spectral data of the four sources and the model
spectra. The observed GeV and TeV data are adapted from Abdo \etal\
(2010) and Aharonian \etal\ (2008), respectively. The model
parameters are given in Table~1. The dashed and dotted lines
represent the model spectra for the upper and lower limits of $\Rc$,
respectively, for both sources A and C (see text in
\S\ref{sec:dif-W28}).} \label{fig:W28N}
\end{figure*}

Now we use the accumulative diffusion model established above to
reproduce the GeV--TeV spectra for the four $\gamma$-ray sources
around SNR W28, showing that the different $\gamma$-ray spectra of
the four sources mainly result from the different spectra of the
runaway protons, as a consequence of the different distances of the
sources (molecular clouds) from the SNR shock from which the
accelerated protons escape.


The current radius of W28 is approximated as $\RSNR\sim12$~pc by
using 2kpc as the distance to it. The current blastwave velocity
$\vSNR$ was inferred to be in the 60--$100\km\ps$ range based on the
optical observations (see Rho \& Borkowski 2002), and hence we adopt
$\vSNR=80\km\ps$. Also following Rho \& Borkowski (2002), we adopt
the intercloud medium density $\no\sim10\cm^{-3}$. For the evolution
in the radiative phase, its age is estimated as $t_{\rm
age}\sim2\RSNR/(7\vSNR)=4.2\E{4}(\RSNR/12\parsec)(\vSNR/80\km\ps)^{-1}
\yr$. The explosion energy is given by
$\ESNR=6.6\times10^{50}(\no/10\cm^3)$
$(\vSNR/80\km\ps)^{1.2}(\RSNR/12\rm pc)^3 \rm erg$. Moreover, we
assume a power-law energy spectrum of the escaping protons of index
$\Gamma=2.2$ (e.g., Giuliani \etal\, 2010), and adopt $\chi=0.1$
(Fujita \etal\ 2009) and $\eta=0.1$ (Blandford \& Eichler 1987).
Therefore, there are three adjustable parameters: $R_{\rm c}$,
$\delta$, and $M_{\rm c}$. These model parameters for the
$\gamma$-ray sources around SNR W28 are shown in Table~1.
 All the spectra of the four sources are well reproduced
(Figure~\ref{fig:W28N}), as described in detail below.
\begin{table}
  \caption{Parameters of the model considered for different sources.}
  \begin{tabular}{lccc|c}
  \hline
   Variable & $R_{\rm c}/\rm pc$ & $\delta$ & $M_{\rm c}/10^4M_{\sun}$
   & $n_{\rm b}/$cm$^{-3}\,\,^{\rm a}$\\
 \hline
 source N & 12 & 0.45 & 4 & $1.6\E{3}$\\
 source B & 22 & 0.35 & 0.2 & $1.1\E{2}$\\
 \hline
 source A$_{\rm min}$ & 100 & 0.6 & 3.2 & $1.8\E{3}$\\
 source A & 140 & 0.65 & 8 & $4.4\E{3}$\\
 source $\rm A_{\rm max}$ & 180 & 0.7 & 13 & $7.2\E{3}$\\
 \hline
 source $\rm C_{\rm min}$ & 55 & 0.55 & 0.64 & $1.0\E{3}$\\
 source C & 100 & 0.6 & 2.4 & $3.9\E{3}$\\
 source $\rm C_{\rm max}$ & 250 & 0.7 & 22 & $3.5\E{4}$\\
 \hline
\end{tabular}
\\$^{\rm a}$ The adoption of the average angular diameter of each
source for deriving the baryon density $n_{\rm b}$ is described in
the text in \S\ref{sec:dif-W28}.
\end{table}

Source N is a well known shock-MC interaction region on the
northeastern boundary of SNR~W28. We adopt $R_{\rm c}\sim R_{\rm
s}(t_{\rm age})\sim12\rm pc$ for this source.
The observed GeV spectrum of source N has a spectral break at
$1.0\pm0.2$~GeV with photon index 2.1 below the break and 2.7 above
the break (Abdo \etal\ 2010). As seen in the top left panel of
Figure~\ref{fig:W28N}, this spectral break together with the TeV
spectrum can be naturally explained by the accumulative diffusion
model without invoking either a breaking power-law proton spectrum
(see Abdo \etal\ 2010) or any artificial low energy cut-off. The
model spectrum peaks at $\sim0.35$~GeV. We note that the TeV
emission with a slope 2.7 implies the same power-law index for the
protons, and this is unlikely to come directly from the shock
accelerated protons (because the typical accelerated particle
spectral slope for acceleration process is 2--2.3) but can be
obtained from a diffusion process.


Source B is located right on the south of SNR~W28 with a projected
distance of $\sim10$~pc from its southern circular boundary
(Aharonian \etal\ 2008), and hence we adopt $R_{\rm c}\sim$22~pc.
In source B and also in source N, which are closely near the SNR,
the contribution from the diffuse Galactic CRs is negligible
compared with that of the diffused protons from the SNR. In our
model, with the increase of radius $R_{\rm c}$ and hence the
increase of the mean diffusion distance, the spectral peak shifts to
higher energies. The peak $\sim0.62$~GeV for source B (see the top
right panel of Figure~\ref{fig:W28N}) is higher than the peak in
source N, which implies the diffusion effect for different diffusion
distances. Source B is also suggested to be associated with HII
region W28A2 which may also possibly explain the $\gamma$-ray
emission, but requires extremely high density of gas (Abdo \etal\
2010).

Sources A and C were not found to have significant counterparts in
the \Fermi\ LAT observation.
 The LAT upper limits together with the \HESS\ data
are indicative of a spectral break in the 10--100~GeV range.
This break is reproduced in Figure~\ref{fig:W28N} (bottom left
panel), in which the concave \gray\ spectra appears as ``M" shape
with two peaks. The underlying proton spectra consist of two
components: the galactic proton background, which is responsible for
the left peak, and the protons coming from SNR W28, which are for
the right one. With the increase of $R_{\rm c}$, the flux density of
CRs accelerated by the SNR is diluted and, compared with the SNR
protons, the relative significance of the diffuse Galactic protons
rises. There are uncertainties in the GeV fluxes for both sources A
and C. The lower limits of $\Rc$ (dotted lines in the bottom left
panel of Figure~\ref{fig:W28N}) are used for the \grays\ produced by
protons escaped from SNR to match the upper limits of the LAT data,
and the upper limits of $\Rc$ (dashed line in the bottom left panel
of Figure~\ref{fig:W28N}) are for the diffuse Galactic protons to
match the LAT upper limits. Meanwhile, with $\Rc$ increased, the
(right) peak of the emission of the escaping SNR protons now shifts
to higher energy than that for source B, thus forming the spectral
break at 10--100GeV.
Although there is less confinement in the GeV realm for source C, we
can determine the upper and lower limits of $\Rc$ of the source (see
the dotted and dashed lines in the bottom right panel of
Figure~\ref{fig:W28N}), in a similar way used for source A.
The fitted values of $\Rc$ for the both sources are much larger than
those of sources N and B (see Table~1), which maybe the basic reason
why the GeV emission for A and C looks faint.

The estimated mass of northeastern cloud (corresponding to the
\gray\ source N) from NANTEN data is $\sim5\E{4}M_{\sun}$ and the
total mass of the southern clouds (including the clumps
corresponding to sources A, B, and C) is $\sim10^5M_{\sun}$
(Aharonian \etal\ 2008). The values of $M_{\rm c}$ fitted for the
four sources in our model (Table~1) seem to be in agreement with the
NANTEN observation. Using the average angular diameters of sources
N, A, B, and C, $18'$, $16'$, $16'$, and $12'$, respectively, as
measured from the outermost contours of the TeV emission in Figure~1
in Aharonian \etal\ (2008), we give the baryon density for each
molecular cloud in Table~1 also.

The reproduction of the four \gray\ spectra produces various values
of parameter $\delta$ in the range of 0.35--0.7. The diffusion
process of CRs results from particle scattering on random MHD waves
and discontinuities. Theoretically, the power-law index of
diffusion, $\delta$, then depends strongly on the spectral energy
density of interstellar turbulence which would originate from fluid
instabilities and magnetic field fluctuations. However, the
mechanisms of the turbulent phenomena are still in hot debate. In
general, three types of isotropic diffusion scalings (Kolmogorov,
$\delta=1/3$; Kraichnan, $\delta=1/2$; and Bohm, $\delta=1$) may
exist in the ISM (see Strong \etal\ 2007). The $\delta$ values
obtained in our model vary just between these values and may arise
from one or a combination of the three types of diffusion, in view
of the fact that the escaping protons diffuse to the four MCs though
a nonuniform and anisotropic interstellar environment around
SNR~W28.





\section{CONCLUSION}\label{sec:conclusion}
An accumulative diffusion model has been established and well
explains the $\gamma$-ray spectra of the four sources around SNR W28
recently obtained by $\HESS$ and \Fermi. The energetic protons
colliding the nearby MCs are considered to be an accumulation of the
diffusive protons escaping from the shock front throughout the
history of the SNR expansion. For the various distances of the MCs
from the SNR centre, the resulted proton spectra can have
prominently different shapes. We have thus fitted the \gray\ spectra
of the four sources and naturally explained the GeV spectral break
of the northeastern source N and the nonsignificant GeV emissions of
the southern sources A and C. The distances of sources A and C from
the SNR centre are found to be much larger than those of sources N
and B. This also implies prospects for inferring the spatial
position of the proton-illuminated MCs around SNRs using the GeV and
TeV spectra.
\grays\ from molecular clouds illuminated by accumulated diffusive
protons from SNR~W28 strongly support the scenario that hadrons
accelerated by SNR shocks contribute to the Galactic CRs.



\section*{Acknowledgments}
Y.C. acknowledges support from NSFC grants 10725312 and the 973
Program grant 2009CB824800.

\bsp

\label{lastpage}

\end{document}